\documentstyle[aps,preprint]{revtex}
\begin{document}
\draft
\title{Solution of scaling quantum networks}
\date{\today}
\author{Yu. Dabaghian and R. Bl\"umel}
\address{Department of Physics, Wesleyan University,
Middletown, CT 06459-0155, USA}
\maketitle

\begin{abstract}
We show that all scaling quantum graphs are
explicitly integrable, i.e. any one of their
spectral eigenvalues $E_n$ is
computable analytically, explicitly,
and individually for any given $n$.
This is surprising, since quantum
graphs are excellent
models of quantum chaos [see, e.g.,
T. Kottos and H. Schanz,
Physica E {\bf 9}, 523 (2001)].
\end{abstract}
\maketitle

Graphs are networks of bonds and
vertices. Figure~1 shows two examples:
a three-bond four-vertex star graph (Fig.~1a) and a
three-bond four-vertex linear graph (Fig.~1b).
A quantum particle moving on the graph turns
the graph into a quantum graph
\cite{QGT1}.
If the quantum particle moves freely on the
graph, subjected only to flux conservation
at its vertices, we call it a {\it standard
quantum graph}. This is the type of quantum graph most
frequently studied in the literature
\cite{QGT1,QGT2,KS,SQA,Kuch}.
A larger class of quantum graphs, including
the standard quantum graphs, are
{\it dressed quantum graphs} \cite{Nova}. A dressed
quantum graph has potentials on its
bonds and $\delta$-functions on its vertices.
The potentials on its bonds are essentially
arbitrary as long as they do not
introduce turning points on the bonds.
But even this case can be dealt with
trivially by re-defining the topology
of the graph.

An important subset of dressed quantum graphs
are {\it scaling quantum graphs} \cite{JETPL,PRL,MF}. In this case
the graph bonds are dressed with scaling
potentials and the graph vertices are
dressed with scaling $\delta$-functions.
A scaling potential is one whose
strength $V_0$ scales with the energy $E$
of the quantum particle according to
$V_0=\lambda E$, where $\lambda$ is a
constant. The strength of a scaling
$\delta$-function scales with $k=\sqrt{E}$.
Scaling potentials and $\delta$-functions
are a natural choice to consider.
On the one hand they frequently occur
in physical systems
\cite{SP1,SP2,SP3,SP4,SP5,SP6},
on the
other hand they are mathematically
convenient, since they allow studying
a quantum system without causing
phase-space metamorphoses \cite{Morph}
in the underlying classical system.
It has been pointed out before \cite{Kohler} that
this is the most natural way of
studying quantum systems, in
particular quantum chaos \cite{Gutz,STOECK}.
Since quantum graphs are popular and
successful models of quantum chaos
\cite{QGT1,QGT2,KS,SQA,Kuch},
it may come as a surprise that
the energy spectrum $E_n$, $n=1,2,\ldots$
of all
scaling quantum graphs is explicitly and
analytically solvable in the form
$E_n=\ldots$, involving only known
quantities on the right-hand side.
In many cases the solutions can be stated
in closed analytical form.

The spectral function $g^{(0)}(k)$ of
a general scaling, dressed quantum graph is
of the form \cite{JETPL}
\begin{equation}
  g^{(0)}(k)=\cos(S_0 k-\pi\gamma_0) -
  \sum_{j=1}^N\, a_j^{(0)}\, \cos(S_j k-\pi\gamma_j),
 \label{1}
\end{equation}
where $S_0>0$ is the total reduced action length of
the graph \cite{JETPL,PRL}, $0<S_j<S_0$ are certain
combinations of the reduced bond actions\cite{JETPL,PRL},
$N$ is the number of action combinations
in (\ref{1}), $\gamma_0$, $\gamma_j$ are
constant phases and $a_j^{(0)}$ are constant
amplitudes. The spectrum $E_n$ of
the quantum graph is obtained by
solving the spectral equation
\begin{equation}
g^{(0)}(k_n^{(0)})=0,\ \ \ n=1,2,\ldots\
\label{2}
\end{equation}
via $E_n=(k_n^{(0)})^2$. For
the purposes of this paper we are
only interested in the positive solutions
of (\ref{2}), and obtain a well-defined
counting index $n$ by defining
$k_1^{(0)}$ to be the first positive root
of (\ref{2}).
As a first step toward the solution
of the general problem,
it was shown in \cite{JETPL,PRL,MF}
that (\ref{2}) can be solved
explicitly in the form $k_n^{(0)}=\ldots$
if the regularity condition
\begin{equation}
\sum_{j=1}^N\, |a_j^{(0)}| < 1
\label{3}
\end{equation}
is fulfilled. In order to substantiate our
claim that (\ref{2}) is solvable
explicitly for {\it all} scaling
quantum graphs, we have to show that
(\ref{2}) is solvable explicitly even
if (\ref{3}) is not fulfilled.

Before we turn our attention to the general
case, we
introduce our methods
with the help of a
simple example.
Let us consider a scaling quantum graph
derived from the three-bond star graph shown
in Fig.~1a by putting the scaling
potentials $V_l(E)=\lambda_l E$,
$0<\lambda_l<1$ on
its three bonds of length $L_l$,
$l=1,2,3$,
require the
``Kirchhoff-type'' \cite{Kuch} flux
conservation condition
$\sum_{l=1}^3 d\psi_l/dx_l=0$
at its central vertex ($\psi_l$ is
the quantum wave function on bond
number $l$ of the graph and
$x_l$ is the coordinate on
bond number $l$) and require
Dirichlet boundary conditions on its three
dead-end vertices. The
spectral equation is of the form (\ref{1})
with $N=3$, $\gamma_1=\gamma_2=\gamma_3=0$
and
\begin{equation}
    S_0=\alpha_1+\alpha_2+\alpha_3,\ \ \
    S_1=-\alpha_1+\alpha_2+\alpha_3,\ \ \
    S_2=\alpha_1-\alpha_2+\alpha_3,\ \ \
    S_3=\alpha_1+\alpha_2-\alpha_3,\ \ \
\label{4a}
\end{equation}
\begin{equation}
 a_1^{(0)}={-\beta_1+\beta_2+\beta_3\over\beta_1+\beta_2+\beta_3},
 \ \ \
 a_2^{(0)}={\beta_1-\beta_2+\beta_3\over\beta_1+\beta_2+\beta_3},
\ \ \
 a_3^{(0)}={\beta_1+\beta_2-\beta_3\over\beta_1+\beta_2+\beta_3},
\label{4}
\end{equation}
where
\begin{equation}
   \alpha_l=\beta_l\, L_l,\ \ \
   \beta_l=\sqrt{1-\lambda_l}, \ \ \
    l=1,2,3.
\label{5}
\end{equation}
The amplitudes in (\ref{4})
do not fulfill the regularity
condition (\ref{3}). In some cases
$\sum_{j=1}^3 |a_j^{(0)}|=1$ (for instance for
$a_j^{(0)}>0$, $j=1,2,3$), and in many cases
$\sum_{j=1}^3 |a_j^{(0)}|>1$,
which strongly violates the regularity
condition (\ref{3}). Since the
methods and techniques presented in
\cite{JETPL,PRL,MF} for obtaining the spectrum
of a graph explicitly
depend crucially on (\ref{3}),
it seems that completely different
methods have to be developed for
general
graphs, such as the three-bond star
graph of Fig.~1a, which do not
fulfill (\ref{3}). There is, however,
a way to reduce (\ref{1}) to a form that
allows to bring the powerful theory of regular
quantum graphs \cite{JETPL,PRL,MF} to bear.
In order to motivate and to illustrate this method,
let us study the case
$\alpha_1=1$, $\alpha_2=7$, $\alpha_3=11$,
$\beta_1=1/10$,
$\beta_2=1/5$, $\beta_3=1/2$.
In this case
$a_1^{(0)}=3/4$, $a_2^{(0)}=1/2$, $a_3^{(0)}=-1/4$,
$S_0=19$, $S_1=17$, $S_2=5$, $S_3=-3$
and the spectral equation is given by
\begin{equation}
g^{(0)}(k) = \cos(19k)-{3\over 4}\cos(17k)-
 {1\over 2}\cos(5k)+{1\over 4}\cos(3k).
\label{6}
\end{equation}
Since
$|a_1^{(0)}|+|a_2^{(0)}|+|a_3^{(0)}|=3/2>1$,
this quantum graph is certainly not regular.
But let us look at the first derivative
of (\ref{6}). Dividing by $S_0$, we
obtain
$$
g^{(1)}(k)=\cos[S_0 k+\pi/2] -
  \sum_{j=1}^3\, a_j^{(1)}\cos[S_j k+\pi/2]=
$$
\begin{equation}
       -\sin(19k)+{51\over 76}\sin(17k)+
         {5\over 38}\sin(5k)
         -{3\over 76}\sin(3k).
\label{7}
\end{equation}
This time we have
$\sum_{j=1}^3 |a_j^{(1)}|=16/19<1$ and
therefore, since (\ref{7}) is precisely of the
form (\ref{1}) and satisfies (\ref{3}), it
can be solved explicitly
using the methods of \cite{JETPL,PRL,MF}.
In particular it was shown in \cite{JETPL,PRL,MF} that
root number $n$ of a spectral equation that
satisfies (\ref{3}), such as
(\ref{7}), is found
in the root interval $[\hat k_{n-1},\hat k_n]$,
where $\hat k_n$ are the root separators
\cite{JETPL,PRL,MF}.
It was also shown in \cite{JETPL,PRL,MF} that the
location of the root separators is entirely
controlled by the local extrema of the
trigonometric function
with the largest action argument.
Thus, in our case, the root separators of
(\ref{7}) are given by
$\hat k_n=(2n+1)\pi/38$.
Since according to \cite{JETPL,PRL,MF} root number $n$
and only root number $n$ is located in the
interval $[\hat k_{n-1},\hat k_n]$, we
can now compute all roots of (\ref{7})
explicitly and individually according to
\begin{equation}
k_n^{(1)}=\int_{\hat k_{n-1}}^{\hat k_n}\,
k\, \left| {dg^{(1)}(k)\over dk}\right|\,
\delta(g^{(1)}(k))\, dk.
\label{8}
\end{equation}
In \cite{FPRE} we show that because of the hermiticity 
of the spectral eigenvalue problem on quantum graphs the
locations of the local extrema of $g^{(0)}(k)$ are separators 
for the roots of $g^{(0)}(k)$. The location of the local 
extrema of $g^{(0)}(k)$, however, are given by the zeros 
of $g^{(1)}(k)$, which, up to
constants, is the derivative of $g^{(0)}(k)$.
Therefore, using the roots $k_n^{(1)}$,
explicitly computed in (\ref{8}),
as the root separators of (\ref{6}),
we obtain, again explicitly and individually,
\begin{equation}
k_n^{(0)}=\int_{k_{n-1}^{(1)}}^{k_n^{(1)}}\,
k\, \left| {dg^{(0)}(k)\over dk}\right|\,
\delta(g^{(0)}(k))\, dk.
\label{9}
\end{equation}
This solves the task of computing the spectrum of our example
of the three-bond dressed star graph explicitly.

In general, given a spectral equation
(\ref{1}) which does not fulfill (\ref{3}),
we generate a chain of derivative spectral equations
$g^{(m)}(k)$, where
$g^{(m)}(k)$ is the $m$'th derivative of (\ref{1})
divided by $S_0^m$,
explicitly given by
\begin{equation}
   g^{(m)}(k)=
   \cos(S_0 k-\pi\gamma_0+m\pi/2) -
   \sum_{j=1}^N\, a_j^{(m)}\,
   \cos(S_jk-\pi\gamma_j
    +m\pi/2),
 \label{10}
\end{equation}
where $a_j^{(m)}=a_j^{(0)}(S_j/S_0)^m$.
Since $S_0<S_j$, there always exists an $M$ such
that the amplitudes $a_j^{(m)}$ satisfy the
regularity condition (\ref{3}), i.e.,
$\sum_{j=1}^N |a_j^{(M)}|<1$.
Therefore, according to \cite{JETPL,PRL,MF},
root separators $\hat k_n^{(M)}$
exist on the level $M$
and the roots $k_n^{(M)}$ of
$g^{(M)}(k)=0$ are explicitly
computable via
\begin{equation}
k_n^{(M)}=\int_{\hat k_{n-1}^{(M)}}^{\hat k_n^{(M)}}\,
k\, \left| {dg^{(M)}(k)\over dk}\right|\,
\delta(g^{(M)}(k))\, dk.
\label{11}
\end{equation}
Since we now know the roots on the level $M$, we
can go one step backwards to level $M-1$.
According to a root-counting argument \cite{FPRE}
based on the Weyl formula \cite{Gutz,STOECK},
the root separators $\hat k_n^{(M-1)}$ on the level 
$M-1$ are the locations of the local extrema of 
$g^{(M-1)}(k)$, which are given
explicitly by the roots $k_n^{(M)}$, which
we know. Therefore, $\hat k_n^{(M-1)}=k_n^{(M)}$
and the roots of
$g^{(M-1)}(k)=0$ can now be computed explicitly,
according to
\begin{equation}
k_n^{(M-1)}=\int_{k_{n-1}^{(M)}}^{k_n^{(M)}}\,
k\, S_0\, \left| g^{(M)}(k)\right|\,
\delta(g^{(M-1)}(k))\, dk.
\label{12}
\end{equation}
Steps (\ref{11}) and (\ref{12})
define a recursive procedure,
\begin{equation}
k_n^{(m-1)}=\int_{k_{n-1}^{(m)}}^{k_n^{(m)}}\,
k\, S_0\, \left| g^{(m)}(k)\right|\,
\delta(g^{(m-1)}(k))\, dk, \ \ \ m=M,M-1,\ldots,2,
\label{13}
\end{equation}
which can be followed until the level 0 is reached and the
roots $k_n^{(0)}$, i.e. the spectrum of the quantum graph, is known
explicitly.

It is important to notice that (\ref{11}) -- (\ref{13}) are
not just formal solutions. They yield
$k_n^{(m)}$, $m=0,\ldots ,M$ explicitly,
by quadratures. Thus (\ref{11}) -- (\ref{13}) constitute
explicit solutions of the problem, very much
in the spirit of the definition of explicit solutions
by quadratures in the theory of differential equations
\cite{DGL}.

Several special cases require discussion.
If the regularity condition (\ref{3}) is fulfilled,
a root $k_n$ lies {\it strictly inside}
of the interval $[\hat k_{n-1},\hat k_n]$.
However, if (\ref{3}) is not
fulfilled, it is possible that a root
$k_n^{(m)}$ coincides with
one of its separators $\hat k_{n-1}^{(m)}$ or
$\hat k_n^{(m)}$.
This is, e.g., the case in our star-graph
example above, where
$k_{18}^{(0)}=\pi$ is a root {\it and} a root
separator of (\ref{6}).
In the parameter space of $\alpha$'s and $\beta$'s
cases like this are extremely rare (nongeneric).
But even if such a case occurs, it does
not present a problem for our theory.
At the contrary, it saves one integration step
since it can always be
checked before performing the integration
in ({\ref{13}), whether one of the
separators $k_{n-1}^{(m)}$ or $k_n^{(m)}$ is
a root of $g^{(m-1)}(k)$. If so, the result
$k_{n-1}^{(m-1)}=k_{n-1}^{(m)}$,
or $k_{n-1}^{(m-1)}=k_n^{(m)}$,
respectively, is known in advance, without actually
performing the integration.

In other special cases the roots of
$g^{(m)}(k)=0$ can be obtained in the
form of
explicit periodic orbit expansions \cite{JETPL,PRL}.
In order to illustrate this, let us return to our
example of the
three-bond star graph. We notice that
the spectral
equation $g^{(1)}(k)=0$ of
the three-bond star graph looks
the same as
the spectral equation \cite{JETPL}
\begin{equation}
g^{(0)}_{\rm 4V-chain}(k)=\sin(S_0 k) + r_2\sin(S_1 k) +
      r_2r_3\sin(S_2 k) - r_3\sin(S_3 k)=0
\label{14}
\end{equation}
of the dressed four-vertex chain graph shown
in Fig.~1b, where
$r_2=(\beta_1-\beta_2)/(\beta_1+\beta_2)$,
$r_3=(\beta_2-\beta_3)/(\beta_2+\beta_3)$
are the reflection coefficients at the
vertices number 2 and 3 of the chain graph,
and the actions $S_0,\ldots,S_3$ are the
same as in (\ref{4a}).
If we arrange for
the bond actions of the chain graph to
equal the
bond actions of the three-bond star graph, and
furthermore arrange for $a_1^{(1)}=-r_2$,
$a_2^{(1)}=-r_2r_3$, $a_3^{(1)}=r_3$,
which is possible if
the scaling constants of the
three-bond star graph fulfill
$\beta_1^2-\beta_2^2+\beta_3^2=0$, then
$g^{(1)}(k)$ of the three-bond star graph is the same
as the spectral equation (\ref{14})
of the associated
four-vertex chain graph and the spectral points
$k_n^{(1)}$ can be stated immediately and
explicitly in the form of
convergent, periodic orbit expansions \cite{JETPL,PRL},
bypassing any integrations that would have been
necessary according to the scheme defined in (\ref{13}).

Although they presented the first examples of
explicitly solvable quantum graphs,
a major shortcoming of \cite{JETPL} and \cite{PRL}
is the fact that the theory presented in
\cite{JETPL} and \cite{PRL} is only applicable
to {\it regular} quantum graphs, i.e. quantum graphs
that fulfill the regularity condition (\ref{3}).
In this paper we showed that the restriction
to regular quantum graphs is not necessary:
all scaling quantum graphs can be solved explicitly.
Nevertheless, the theory presented in \cite{JETPL,PRL,MF}
provides an indispensable foundation without which
the present theory would not be possible.

A conceptual advance is the following. Frequently
an operational definition of quantum chaos, or
a quantum chaotic regime, is the ``loss of quantum
numbers''. To illustrate, let us consider a
Hamiltonian system with Hamiltonian $\hat H=\hat H_0
+\mu \hat V$, where $\hat H_0$ is an integrable
Hamiltonian, $\mu$ is a real parameter and
$\hat V$, with respect to and in conjunction with
$\hat H_0$, is a
nonintegrable perturbation.
Many quantum systems,
for instance the hydrogen atom in a strong
magnetic field \cite{SP1}, can be described in
this way. For $\mu=0$ the system is integrable
and possesses a complete set of quantum numbers

that can be obtained, at least approximately,
using EBK quantization \cite{Gutz,STOECK}.
As the parameter $\mu$ increases, EBK
quantization breaks down
and the system
makes a transition to quantum chaos.
This explains the frequently employed practice
of characterizing the onset of quantum chaos by
a loss of quantum numbers,
since the breakdown of the EBK quantization scheme
implies the loss of quantum numbers.
The results obtained in this paper, however, show
that this is not necessarily a good way to
characterize quantum chaos. Although
not strictly chaotic in the classical limit
(due to ray-splitting \cite{SP2,SP3,Couch,RS2} the term
{\it stochastic} may characterize the situation
better),
quantum graphs were shown by many authors
\cite{QGT1,QGT2,KS,SQA,Kuch}
to be excellent models of quantum chaos.
Yet, our results above show that a well-defined
quantum number, the counting index $n$, still
exists, and produces explicit energy levels in
exactly the same spirit as the EBK quantization
scheme.

The iteration scheme (\ref{13}) is perhaps the
most interesting feature of our method of explicitly
solving quantum graphs. We call the smallest $M$
that ``regularizes'' a given quantum graph
(i.e. the amplitudes of $g^{(M)}$ fulfill (\ref{3})),
the {\it order} of the quantum graph. For any given
quantum graph its order is unique. Since the order
$M$ of a quantum graph determines the length of
the bootstrapping iteration scheme (\ref{13}), it is
possible that the order of a quantum graph is also
an indication of the complexity of its spectrum.
Quantum iterations similar to (\ref{13}) were
studied before \cite{SPRL} and were found to
lead to sensitivity and chaos on the quantum level.
This may explain the reason why certain quantum graphs
are such good models of quantum chaos
\cite{QGT1,QGT2,KS,SQA,Kuch}
and the order $M$ of the quantum graph may be
an indication of how well a given quantum graph
can be described in terms of the usual diagnostic
tools of quantum chaos, such as, e.g.,
random matrix theory
\cite{Gutz,STOECK,Haake}.

The authors acknowledge financial support by the
National Science Foundation under Grant No. 9984075.

\pagebreak

\centerline{\bf Figure Captions}

\bigskip \noindent
{\bf Fig.~1:} (a) Dressed three-bond star graph and (b)
              dressed four-vertex chain graph.
              Different potential strengths on the bonds
              are indicated by different thickness of the bonds.
              Different vertex strengths are indicated by
              different dot sizes representing the vertices.

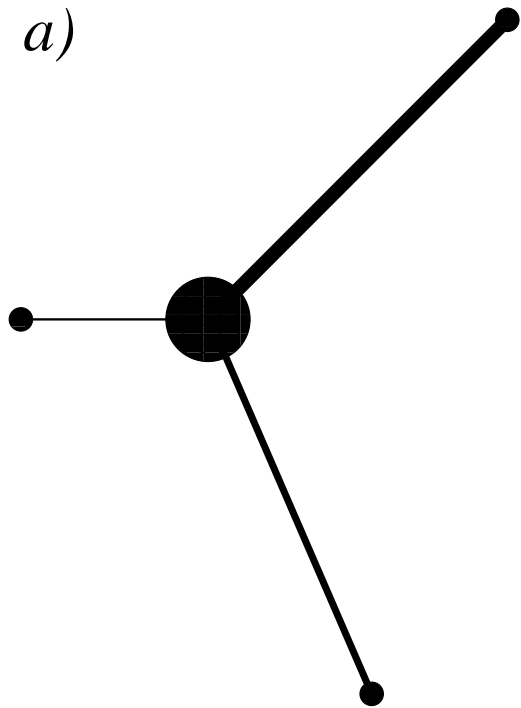
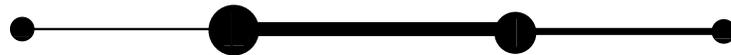
\end{document}